\documentstyle[11pt,newpasp]{article}

\begin{document}

\title{Extragalactic planetary nebulae as mass tracers: biases in the estimate 
of dynamical quantities}

\author{M. Arnaboldi,}
\affil{Osservatorio Astronomico di Capodimonte, Naples, Italy}
\author{N. Napolitano, M. Capaccioli}
\affil{Osservatorio Astronomico di Capodimonte, Naples, Italy}
\affil{Universit\`{a} ``Federico II'', Naples, Italy}




\begin{abstract}
Planetary nebulae (PNe) are very important kinematical tracers of the outer 
regions of early-type galaxies, where the integrated light techniques fail. 
Under {\em ad hoc} assumptions, they allow measurements of rotation velocity 
and velocity dispersion profile from descrete radial velocity fields. 
We present the results on the precision allowed by different set of 
radial velocity samples, discuss the hypotheses in the 
analysis of descrete velocity fields and their impact on the inferred
kinematics of the stellar population.
\end{abstract}


\section{Introduction}
With 4m telescopes and multi-object and/or multifiber spectrographs it has 
been possible to obtain information on the kinematics of the outer regions of 
ellipticals by measuring the radial velocities of individual stars
during their phase of planetary nebulae. This is crucial because 
PNe are found at very large distances from galaxy centers where kinematical
measurements based on standard integrated stellar light techniques 
are no longer possible.
These radial velocity samples obtained from the PNe 5007 \AA [OIII] emission 
in giant Es and S0s (D=15-17 Mpc) contain up to 50 PNe radial 
velocities (Arnaboldi et al. 1998), and larger samples are available only 
for nearer objects (i.e. NGC 5128, Sombrero, NGC 3115). 
Therefore there might be biases introduced by small number statistics that
need to be investigated and understood.\\
Based on a statistical approach, we investigate the possibility of 
re-building the actual kinematics of spherical systems starting from descrete 
radial velocity fields.
We build equilibrium systems for which dynamical and kinematical parameters 
are known (spherical model + known velocity field).
By Montecarlo simulations we produce 100 ``observational sets'', each for
a given sample size, i.e. 50, 150 or 500 randomly chosen stars,  
and then analyse each sample to determine the rotation curves and 
velocity dispersion profiles.
We account also for the measuring errors ($\sim$30\% of the maximum
rotational velocity).
The aim of this work is to address the following questions: what is it the 
precision allowed 
by the different statistical samples in  determining the kinematical 
quantities? Furthermore, may the hypotheses on the internal rotational 
structure introduce any biases?

\section{Model procedure}
We assume a constant M/L ratio and the Hernquist model (Hernquist 1990) for 
the dimensionless mass density distribution 
\begin{equation}
\rho(r) = N \frac{1}{2\pi r (1+r)^3}
\label{rho}
\end{equation}
where $r$ is in units of a core radius $a$ and $N$ is a normalisation constant 
chosen as to have a total mass M$_t$=1 within a distance of 18$a$ from the 
center ($\sim 10 R_e$). The cumulative mass distribution is
\begin{equation}
M_l(x)=4\pi \int_0 ^x x'^2 \rho(x')dx'
\label{mascu}
\end{equation}
In spherical models with dark matter, we consider an additional mass 
contribution coming from an isothermal halo, and can write the dark matter 
density as (Dubinsky \& Carlberg 1991):
\begin{equation}
\rho_d(r)= \frac{M_d}{2\pi}\frac{d}{r}\frac{1}{(r+d)^{3}}
\label{rhdark}
\end{equation}
where $M_d$, $d$ and $r$ have the same meaning as in eq.~(\ref{rho}).
In eq.~(\ref{rhdark}), we take $d=10a$ and $M_d=7.7.M_l$. 
The cumulative mass distribution is defined as in eq.~(\ref{mascu}) 
with total mass:
\begin{equation}
M_t(r)=M_l(r)+M_d(r).
\label{mast}
\end{equation}
The model for the rotational velocity is
$$
V_{rot}(x)=\frac{V_0x}{\sqrt{r_0^2+x^2}}
$$
where $V_0$ is the maximum rotational velocity, $r_0$ is a scale factor and 
$x$ is the distance from the rotational axis (cylindrical rotation).\\
The PNe distribution is computed taking into account the selection effects 
of the bright continuum background from the stars in the central region 
of the galaxy. This selects a projected radius $R_{lim}$ onto
the sky plane out of which the PNe sample is complete.

\section{The descrete radial velocity field}
From the mass density in space, $\rho(r)$, we extract via 
Montecarlo a star at position $(x_p, y_p, z_p )\,=\, (r_p,\theta_p,
\phi_p)$. We check if $R_p^2 = X_p^2 + Y_p^2$, where 
($X_p,Y_p$) are the star projected coordinate onto the Sky Plane, 
is greater that the $R_{lim}$, the completeness limit radius.
In the 3D space, we assign to the selected star its velocity vector 
under the hypothesis of ``{\em local isothermal approximation}'', i.e. 
the observed distribution of radial velocity profiles in Es are 
Gaussian with maximum deviation of 10\% (Bender et al. 1994). 
This is consistent with taking F(r,{\bf v}) as product of three Gaussians.
The $\sigma_{r}, \sigma_{\theta}, \sigma_{\phi}$ components are
derived from the Jeans equation with the adopted $\rho(r)$ for the mass 
distribution, assuming isotropy for the velocity ellipsoid and the adopted 
rotation curve.\\
Once the 3D velocity vector is computed, the radial velocity is derived
by projection of the velocity vector along the line-of-sight.
This procedure is iterated for all the stars in a given sample 
and the 2-D descrete velocity field $V_{obs}(X,Y)$ is derived; a 500 PNe
sample is shown in fig.1.
\begin{figure*}
\includegraphics{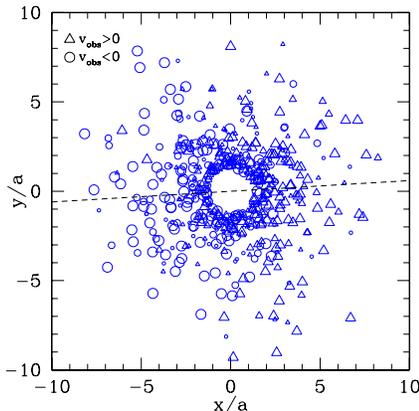}
\vspace{5.3cm}
\caption{A simulated radial velocity field: the dotted line indicates
the direction of maximum gradient.}
\end{figure*}

\section{The analysis procedure}
For each given V$_{obs}(X,Y)$ sample, we perform a simple analysis 
of the velocity field as follows:
we fit a rigid rotator (bilinear function) 
$
v_{rad}=a+bX+cY
$
and the results of this fit are compared with the results of a flat 
rotational curve fit (flat-fit) of the form
$
v_{rad}=v_{sys}+v_{0}\cos(\varphi-\varphi_{0})
$
then we obtain the field of the residuals for each of the two 
interpolated fields
$
\Delta v= V_{obs}-v_{rad}
$.\\
The aim of this procedure is to determine the characteristics of the 
velocity field: systemic velocity, direction of the maximum velocity 
gradient (Z1), velocity gradient and maximum rotational velocity.
Along $Z1$, we select a slice of width $dZ1$ in order to trace the 
kinematics along this relevant axis; $dZ1$ must be small enough to trace 
the kinematics related to Z1 and large enough to allow a significant 
statistical sample. To describe the velocity dispersion profile, we analyse 
the residual fields from the two adopted fits, by computing averages along Z1 
in bins, which contains at least 10 points.\\
As a check for the presence of biases, along the direction Z1 of maximum 
gradient, we obtain the rotation curve and the velocity dispersion 
profile simply as the average and radial velocity RMS in each of the 
given bins used in the analysis of the residuals, and we refer to this 
as the ``No-fit'' procedure.

\section{Results and conclusions}
As a first result we obtained that the direction of maximum gradient is 
independent of the rotational structure adopted for fit.
For simulated samples with 500 PNe, the comparison between the 
simulated results with the expected model values show that the
bilinear fit may introduce {\bf an over estimate} of the velocity dispersion 
in the outer bins, depending on the {\bf intrinsic rotational structure} 
of the galaxy. This bias is a function of the $v/\sigma$ ratio as shown in 
fig.~2. The no-fit procedure does not introduce biases in the 
determination of the velocity dispersion profile.
\begin{figure*}
\includegraphics{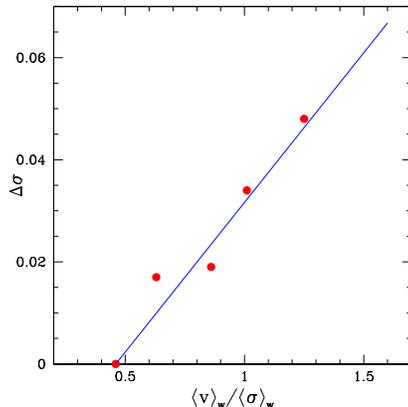}
\vspace{5.3cm}
\caption{The grey points are the values of the difference between the
velocity dispersion determined in the last bin along Z1 and the 
expected value (weight average). The continuous line indicates the 
linear regression.
The behaviour showed by the residuals indicates that higher rotation in 
the model causes the bilinear fit to introduce a larger bias in the
velocity dispersion determination.}
\end{figure*}
For samples with 50 PNe, in the outer bins we over-estimate the velocity 
dispersion by an amount which is dependent from the {\bf underlying mass 
distribution}. If we take into account the weight effects in the binning, 
using the surface brightness distribution (i.e. surface density distribution 
via M/L ratio), we found a better consistency with the simulated data. The 
precisions obtained for the kinematical quantities depend on the 
selection performed along Z1. Our best mean estimate is 15\% for 500 PNe 
and 22\% for 50 PNe.\\
PNe can be used very efficiently as kinematical probes to trace the dynamics 
of the outer stellar halos in giant early-type galaxies provided that
the analysis of the descrete radial velocity fields avoids any systematics
depending on the assumption of the internal angular momentum distribution.

%

\end{document}